\documentclass[12pt]{iopart}

\usepackage{graphicx}
\usepackage{hyperref}
\begin{document}

\title[Negative parity vector-coupling schemes]{Seniority-two valence-shell building blocks of the octupole phonon}

\author{
M~\textsc{Scheck},$^{1,2}$
R~\textsc{Chapman},$^{1,2}$
K~\textsc{Mashtakov},$^{1,2}$\footnote[1]{Present Address: Department of Physics, Guelph University, Guelph, Ontario N1G2W1, Canada}
R~\textsc{Meeten},$^{1,2}$
P~L~\textsc{Sassarini},$^{1,2}$ and
P~\textsc{Spagnoletti}$^{1,2}$\footnote[2]{Present Address: Department of Physics, University of Liverpool, Liverpool L69 7ZE, UK}\\
\address{
$^{1}$School of Computing, Engineering, and Physical Sciences, University of the West of Scotland, Paisley PA1 2BE, UK \\
$^2$SUPA, Scottish Universities Physics Alliance, UK \\}
}

\ead{marcus.scheck@uws.ac.uk}

\begin{abstract}
The relative ordering of $J^-$ levels of multiplets resulting from two-body excitations, which include a $J^{\pi}=3^-$ state that can contribute to the octupole phonon, are investigated in a simplistic shell-model approach. To calculate the relative level ordering, harmonic oscillator wavefunctions and a residual $\delta$ interaction are used. The simplistic approach confirms for the particle-particle channel, the often stated preference of the $\Delta j=3, \Delta l =3$ subshell combination over the $\Delta j=3, \Delta l =1$ subshell combination through an enhanced energy gain. Furthermore, it is shown that, in the particle-hole channel, the gain is less pronounced for the $\Delta j=3, \Delta l =3$ subshell configuration. In combination with the overall structure of an oscillator shell, these results explain the comparatively low excitation energy for 3$^-_1$ excitations observed for the octupole-soft proton and neutron numbers predicted by various models.
\end{abstract}

\maketitle

\section{Introduction}

\begin{figure}[ht]
\centering
\includegraphics[width=0.7\textwidth, viewport=1 1 482 176, clip=true]{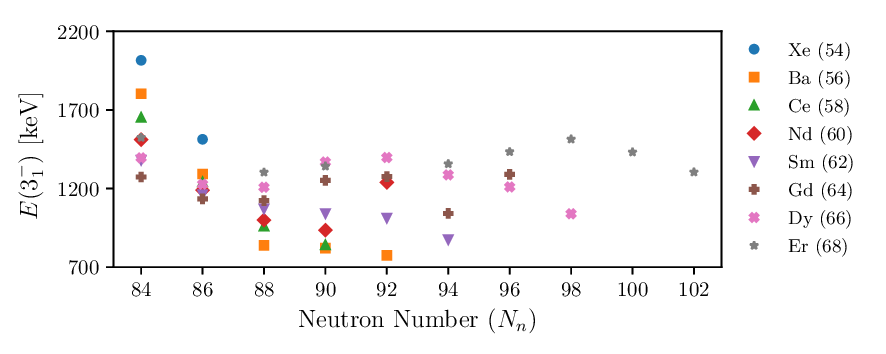}
\caption{Excitation energies of the lowest-lying $3^-_1$ levels in nuclei northeast of $^{132}$Sn. The proton number $Z$ is provided in the legend behind the symbol of the element. Data are taken from Ref.~\cite{Kibedi}.}
\label{figure_1}
\end{figure}

\begin{figure}[ht]
\centering
\includegraphics[width=0.9\textwidth, viewport=1 1 493 176, clip=true]{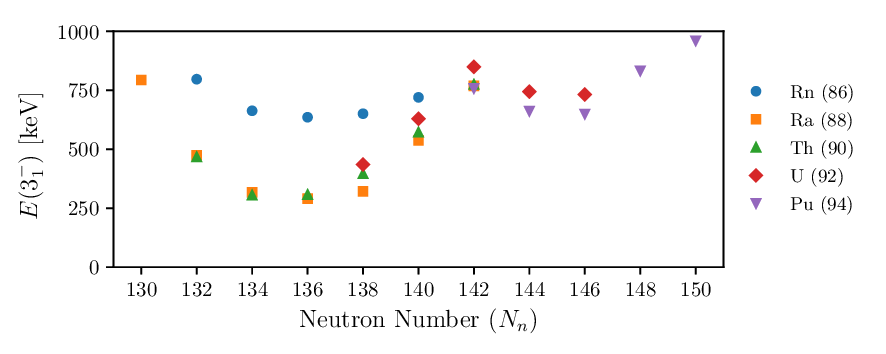}
\caption{Excitation energies of the lowest-lying $3^-_1$ levels in nuclei northeast of $^{208}$Pb. The proton number $Z$ is provided in the legend behind the symbol of the element. Data are taken from Ref.~\cite{Kibedi}.}
\label{figure_2}
\end{figure}

\begin{figure}[ht]
\centering
\includegraphics[width=0.7\textwidth, viewport=1 1 475 176, clip=true]{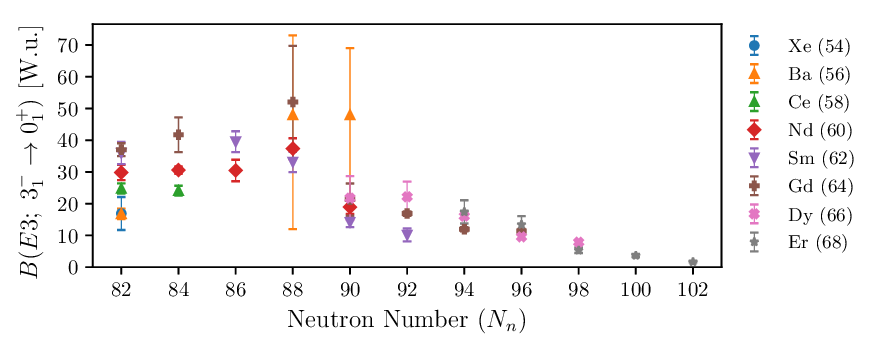}
\caption{$B(E3, 3^-_1 \rightarrow 0^+)$ reduced transition strengths of the lowest-lying $3^-_1$ levels in nuclei northeast of $^{132}$Sn. The proton number $Z$ is provided in the legend behind the symbol of the element. Data are taken from Refs.~\cite{Kibedi,Bucher1,Bucher2}.}
\label{figure_3}
\end{figure}

\begin{figure}[ht]
\centering
\includegraphics[width=0.7\textwidth, viewport=1 1 420 176, clip=true]{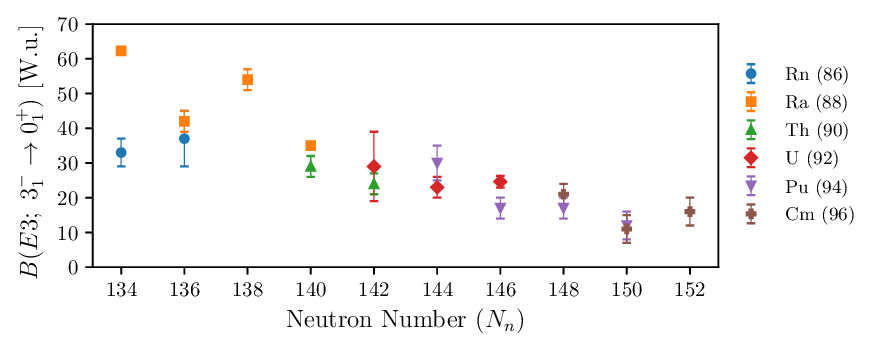}
\caption{$B(E3, 3^-_1 \rightarrow 0^+)$ reduced transition strengths of the lowest-lying $3^-_1$ levels in nuclei northeast of $^{208}$Pb. The proton number $Z$ is provided in the legend behind the symbol of the element. Data are taken from Refs.~\cite{Kibedi,Liam,Peter1,Peter2,Peter3,Pietro}.}
\label{figure_4}
\end{figure}

Nuclei situated in the Segre chart near the proton numbers $Z =34, 56, 88$ and neutron numbers  $N_{n} = 34, 56, 88, 134$ are predicted to exhibit a pronounced octupole collectivity \cite{Butler}. An enhanced collectivity results in a low excitation energy of the first excited state of the given spin and parity and a high excitation probability to this level from the ground state. Indeed, as shown in Figures~\ref{figure_1} and \ref{figure_2}, the excitation energies of the first excited $3^-$ levels for the $Z=56$, $N_n\approx 88$ barium nuclei and the $Z=88$, $N_n \approx 134$ radium are lowest in their respective mass region \cite{Kibedi}. The presently available data for the $B(E3, 0^+ \rightarrow 3^{-}_{1})$ excitation strength for the $Z=56$, $N_{n}=88$ mass region are shown in Figure~\ref{figure_3}. Apart from $^{144}$Ba \cite{Bucher1} and $^{146}$Ba \cite{Bucher2}, near the octupole soft numbers, the trend indicates stronger $B(E3, 0^+ \rightarrow 3^-_1)$ values at higher proton $Z$ numbers, peaking for the gadolinium isotopes, e.g.\@ see Ref.~\cite{PASCU}. In fact, the $B(E3)$ value of $^{152}$Gd (Z= 64) is one of largest observed so far, but it has a comparatively large uncertainy. As shown in Figure~\ref{figure_4}, for the $A \approx 222$ ($Z=88, N_n= 134$) region, only a few $B(E3, 0^+ \rightarrow 3^-_1)$ values are known, mostly for radon and radium nuclei \cite{Kibedi,Liam,Peter1,Peter2,Peter3,Pietro}, but not for the thorium, uranium, and plutonium isotopes near $N_{n} = 134$. As calculated in Ref.~\cite{Luis1} and shown in Fig.~19 of Ref.~\cite{Peter4}, the $B(E3, 0^+ \rightarrow 3^-_1)$ strength in these nuclei is predicted to increase with $Z$ beyond the octupole soft number of $Z=88$. Fig.~1 in Ref.~\cite{Scheck1} shows an inverse sum-rule $B(E3, 3^-_1 \rightarrow 0^+)/E_{3^-_1}$ value, which renormalises the $B(E3, 3^- \rightarrow 0^+)$ strength to the $E_{3^-_1}$ excitation energy. This plot, combining the two most significant observables for octupole collectivity, demonstrates the special character of the $Z=56$ barium isotopes near $N_n=88$ and $Z=88$ radium isotopes near $N_n=134$.

In a simplistic approach, which considers exclusively senority-two valence-shell excitations, only a comparatively low number of configurations can contribute to the $3^-_1$ octupole phonon. Of course, this approach neglects excitations with higher seniority and cross-oscillator shell contributions to the collective wavefunction. Despite having only small amplitudes, the latter can as sum significantly contribute to the transition probabilities, a fact that has recently been shown for the quadrupole degree of freedom, e.g.\@ see Ref.~\cite{Walz}, and is also evidenced by the strong $B(E3, 0^+ \rightarrow 3^-_1)$ values of the stable lead isotopes \cite{Kibedi, Henderson}. While the excitation probabilities are subject to the subtle interplay and interference effects between various contributions to the wavefunction, the excitation energy can be expected to be less affected by these cross-oscillator shell contributions. Compared to the quadrupole degree of freedom, the octupole collectivity is low and the energies of these low collective excitations can still be expected to be influenced by the positioning of the lowest-lying particle-particle/particle-hole component. Hence, it is interesting to investigate the behaviour of the valence-shell two-particle or one-particle one-hole excitations that contribute to the collective wavefunction of the $3^-_1$ octupole phonon. 

\begin{figure}[ht]
\centering
\includegraphics[width=0.7\textwidth, viewport=3 0 784 473, clip=true]{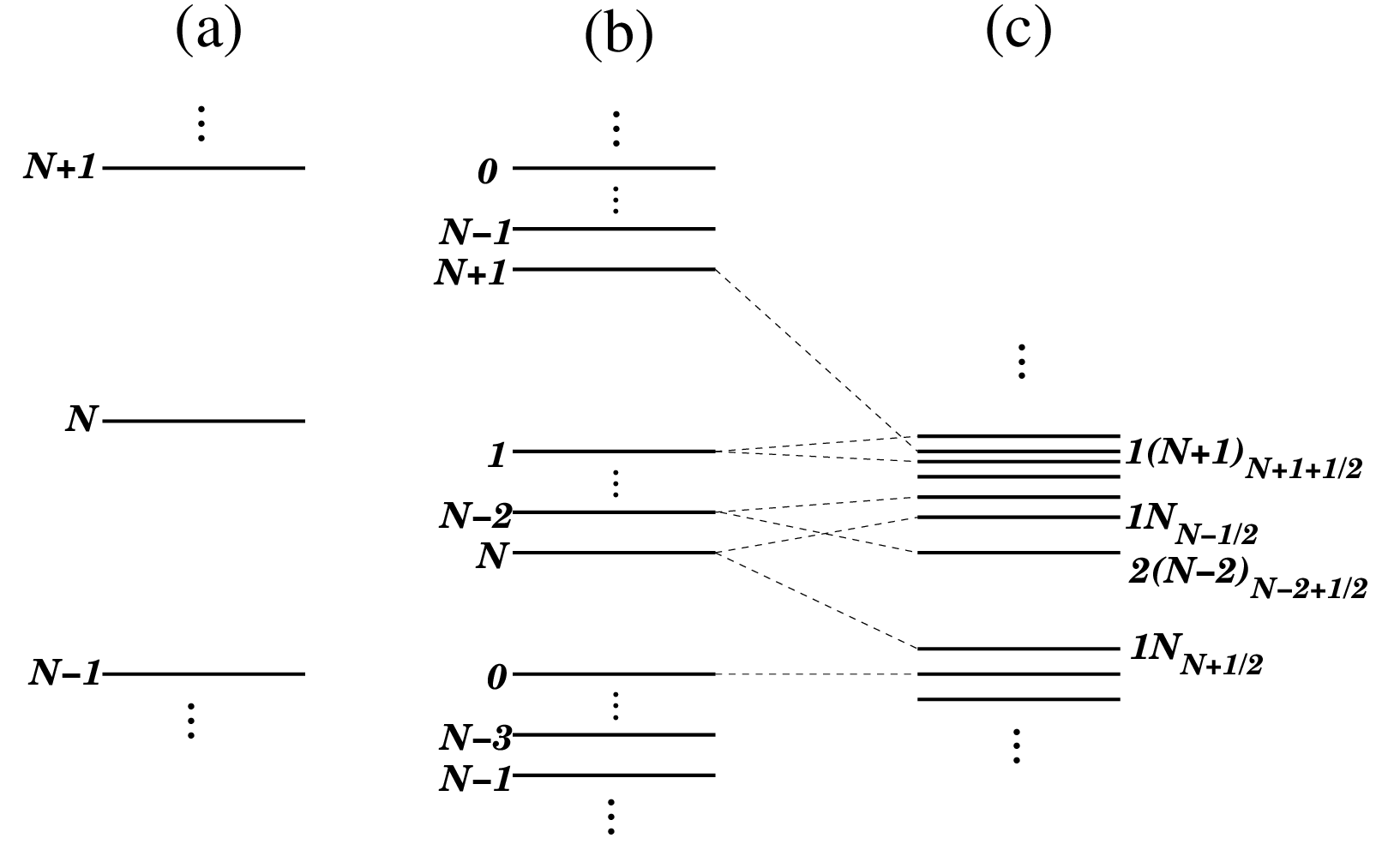}
\caption{Schematic representation of the nuclear shell structure as relevant to this contribution. Part~(a) is in the harmonic oscillator stage and shells are labeled by the oscillator quantum number $N$. In part~(b) a $l^2$ term  is introduced, which flattens the central part of the HO potential, analogously to the Woods-Saxon potential, and lifts the l degeneracy. The subshells are labelled with the orbital angular momentum $l$, which is expressed in terms of the oscillator shell quantum number $l= N, N-2, N-4, ..., 0$ or $1$. In part~(c) the spin-orbit force is considered. The relevant subshells for the $\Delta j=3, \Delta l=3$ and $\Delta j=3, \Delta l=1$ combinations are labeled with $nl_j$, where $n$ is the principal quantum number, $l$ the orbital angular momentum quantum number, and $j$ the quantum number of the total angular momentum $j=l\pm 1/2$. The quantum numbers $l$ and $j$ are expressed using the quantum number of the oscillator shell $N$. For a discussion see text.}
\label{figure_5}
\end{figure}

The negative parity of the low-lying $3^-$ excitation facilitates the necessity to include the unique-parity intruder subshell in the bilinear form of the $[nl_j, n^{\prime}l^{\prime}_{j^{\prime}}]_{J^-}$ particle-particle or $[nl_j, (n^{\prime}l^{\prime}_{j^{\prime}})^{-1}]_{J^-}$ particle-hole configuration. For a given oscillator shell, labeled by the oscillator quantum number $N$, the orbital angular momentum $l$ of the individual subshells can adopt $l= N, N-2, N-4, ..., 0$ or $1$ and all subshells have the parity $\pi = (-1)^N$. The subshells are labelled by $nl_j$, where $j =l \pm s = l \pm 1/2$ is the coupled angular momentum composed of orbital $l$ and spin $s=1/2$ parts and $n$ represents the principal quantum number\footnote[1]{In this contribution the nomenclature $n=k+1=1,2,3,...$ of Talmi and deShalit is used, for which the principal quantum number includes the root at the origin.} that corresponds to the roots of the radial part of the wave function. As shown in Fig.~\ref{figure_5}, the subshells can all be uniquely labeled with the oscillator quantum number $N$. The intruder subshell $l=(N+1)$ does have the total angular momentum $j= l+ s = l+ 1/2=N+1+1/2$ and parity $\pi = (-1)^{N+1}$. Hence, the unique parity subshell can be parameterised as $nl_j = 1(N+1)_{N+1+1/2} = 1(N+1)_{N+3/2}$. Since the $nl_{j} = 1N_{N+1/2}$ subshell with the highest $l=N$ in the $N$-th oscillator shell itself has become the extruder orbital to the $N-1$ oscillator shell, only its spin-orbit partner $nl_j = 1N_{N-1/2}$ and the  $nl_j =2(N-2)_{N-2+1/2}= 2(N-2)_{N-3/2}$ subshells can couple with the intruder subshell to form a $3^-$ excitation. Consequently, for both, protons or neutrons, only the two subshell configurations with $\Delta j=3; \Delta l=1$: $[1(N+1)_{N+3/2}, 1N_{N-1/2}]_{3^-}$ and $\Delta j=3; \Delta l=3$: $[1(N+1)_{N+3/2}, 2(N-2)_{N-3/2}]_{3^-}$ contribute to the octupole phonon. Examples of these particular configurations are given in Table~\ref{table_1}. The $\Delta j =3, \Delta l =3$ configurations are predicted to be most susceptible to the residual long-range octupole-octupole term of the multipole expansion of the nuclear force \cite{Butler}.


The gross structure of an oscillator shell (see Figure~\ref{figure_5}), which is labeled by its oscillator quantum number $N$, is dominated by two effects. First, the centrifugal $l^2$ dependence lowers the energy of high-$l$ orbits. For the oscillator potential this term mimicks the flat central part of the Woods-Saxon potential. As a consequence, the $nl_j= 1N_{N-1/2}$ subshell, even after spin-orbit coupling has been considered, is found amongst the lowest-lying subshells of an oscillator shell. Second, the spin-orbit interaction between the $nl_j = 2(N-2)_{N-2+1/2}$ subshell and its spin-orbit partner $nl_j = 2(N-2)_{N-2-1/2}$ results in a lowering of the $2(N-2)_{N-2+1/2}$ subshell, which is then found at the low-energy end of the $N$-th oscillator shell. In combination with blocking of configurations, when filling the unique parity subshell, these facts explain why nuclei with strong octupole correlations are found in the chart of nuclei north-east near doubly-magic nuclei. In these mass regions, the Fermi levels for protons as well as for neutrons are situated in the $nl_j = 1N_{N-1/2}$ or $2(N-2)_{N-2+1/2}$ subshells.

\begin{table}[b]
\caption{\label{table_1} Valence-shell subshell configurations for likewise particles that can contribute to the $J^{\pi}=3^-$ octupole excitation within an oscillator shell labeled by the oscillator quantum number $N$.}
\begin{indented}
\item[]\begin{tabular}{@{}ccc}
\br
$N$ & $\Delta l =3, \Delta j=3$ & $\Delta l =1, \Delta j=3$  \\
\mr
3 & $[2p_{3/2^-},1g_{9/2^+}]$ &  $[1f_{5/2^-},1g_{9/2^+}]$\\
&&\\
4 & $[2d_{5/2+},1h_{11/2^-}]$ & $[1g_{7/2^+},1h_{11/2^-}]$\\
&&\\
5 & $[2f_{7/2^-},1i_{13/2^+}]$ & $[1h_{9/2^-},1i_{13/2^+}]$\\
&&\\
6 & $[2g_{9/2+},1j_{15/2^-}]$ & $[1i_{11/2^+},1j_{15/2^-}]$\\
\br
\end{tabular}
\end{indented}
\end{table}

Interestingly, in a naive picture, one would expect the excitation energy of the octupole phonon to be lowest when the number of participating nucleons is maximum and when both subshells of the $N$-th oscillator shell, allowing a $\Delta j = 3$ configuration, are completely filled. For example, in the $N=5$ oscillator shell, where the $2f_{7/2^-}$ and $1h_{9/2^-}$ subshells are present, the octupole softest neutron number should be $N_{n} = 82 + \sum_{j} (2j+1)= 82+ 8 +10 =100$, and not $88$. Beyond $N=100$, the filling of the unique-parity subshell and the resulting blocking of particle excitations reduces the collectivity. Indeed, as mentioned above, near $A\approx 144$ the $B(E3)$ values somehow support for the proton numbers an enhanced collectivity beyond the octupole soft number (see Figure~\ref{figure_3}). However, the excitation energy of the $3^-_1$ levels is lowest for the protons the octupole soft numbers, which are interestingly two nucleons short of the fully filled $nl_j = 2(N-2)_{N-3/2}$ subshell that forms with the unique parity subshell a $\Delta j=3; \Delta l=3$ combination. For completness it must be mentioned that for many cases the number of both types of valence nucleons is already considerable and the nucleus will adopt a quadrupole deformed shape. Consequently, the spherical approach used in this contribution should be treated with care. Nevertheless, is it interesting to investigate if the level ordering of the possible $J^-$ levels resulting from the $[nl_j, n^{\prime}l^{\prime}_{j^{\prime}}]_{J^-}$ coupling of the two subshell combinations is strongly affected by the two involved angular momenta and by whether one considers the particle-particle or the particle-hole channel.  

\begin{figure}[ht]
\centering
\includegraphics[width=0.7\textwidth, viewport=0 0 400 215, clip=true]{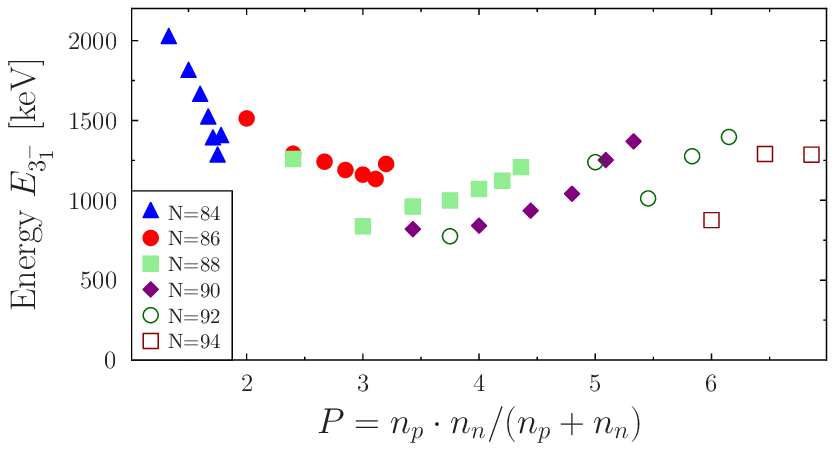}
\caption{Excitation energy $E_{3^-_1}$ of the first $3^-_1$ state as function of the $P$ factor \cite{Casten1,Casten2,Casten3}, which represents a measure for the number of proton-neutron interactions of the valence nucleons renormalised by the number of valence nucleons. The data contain the available information in the $A \approx 144$ mass region.}
\label{figure_6a}
\end{figure}

Beyond this approach for configurations of likewise particles, one must bear in mind that, for an open-shell nucleus, in which proton as well as neutron excitations contribute to the complex wavefunction of the collective state, the proton-neutron interaction between the protons in a given subshell and the neutrons in another subshell does have an effect on the excitation energy. Interestingly, with respect to the proton-neutron interaction, the nuclei in which octupole soft numbers are realised for both protons and neutrons can be classified according to the occupied oscillator shells. This involves nuclei near the $N_{n}=Z$ line, e.g.\@ $^{68}$Se and $^{112}$Ba, in which protons $(\pi)$ and neutrons $(\nu)$ occupy orbitals in the same oscillator shell $(N_{\nu} = N_{\pi} = N)$, and nuclei on the neutron-rich side of the valley of stability. For the latter nuclei, for example $^{144}$Ba and $^{222}$Ra, the neutron oscillator shell quantum number $N_{\nu}$ equals the proton oscillator shell quantum number $N_{\pi}$ incremented by one: $N_{\nu} =N_{\pi}+1$. As a consequence, the proton-neutron couplings are different for either of these realisations. The possible proton and neutron couplings for the $N_{\nu}= N_{\pi}$ case are listed in Table~\ref{table_pnNZ}, and Table~\ref{table_pnN1Z} presents selected subshell combinations for the $N_{\nu} = N_{\pi}+1$ case. For the latter, only combinations are given for which the two involved subshells are represented by one of the previously named subshells that can couple to a 3$^-$ level for the proton-proton or neutron-neutron channels as well. Besides these, in the proton-neutron channel, a multitude of other subshell combinations can also couple to a 3$^-$ state. 

An interesting aspect of this difference in orbitals of protons and neutrons between the $N_{\nu} = N_{\pi}$ and $N_{\nu} = N_{\pi}+1$ cases is the electric dipole ($E1$) moment associated with quadrupole-octupole coupling (see Refs.~\cite{Scheck1,Kneissl1,Andrejtscheff} and references therein). It will be interesting to investigate, whether the occupancy of identical orbitals for $N_{\nu} \approx N_{\pi}$ nuclei results in a vanishing separation of proton and neutron fluids and, therefore, the associated $E1$ moments and, furthermore, how these $E1$ moments compare to those of nuclei in the $N_{\nu} = N_{\pi}+1$ neutron-rich mass regions. 

\begin{figure}[ht]
\centering
\includegraphics[width=0.7\textwidth, viewport=0 0 400 215, clip=true]{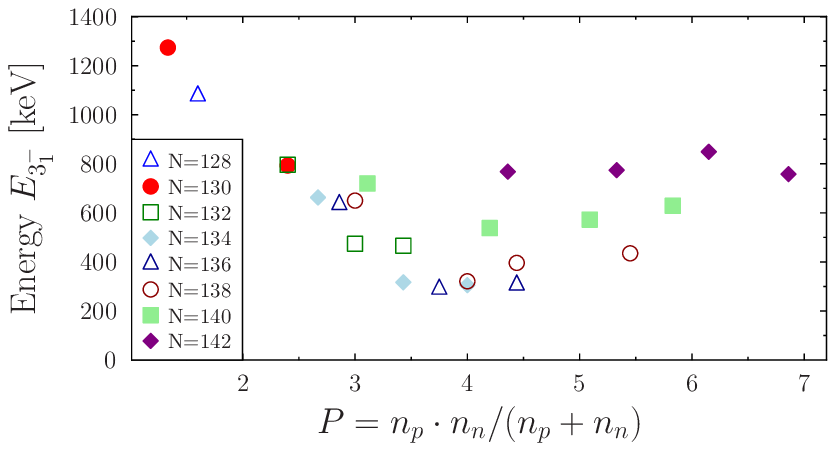}
\caption{Excitation energy $E_{3^-_1}$ of the first $3^-_1$ state as function of the $P$ factor \cite{Casten1,Casten2,Casten3}, that represents the number of proton-neutron interactions of the valence nucleons renormalised by the number of valence nucleons. The data contain the available information in the $A \approx 222$ mass region.}
\label{figure_6b}
\end{figure}

\begin{table}[b]
\caption{\label{table_pnNZ} Proton-neutron $[\rho_1 (n_1 l_1)_{j_1}, \rho_2(n_2 l_2)_{j_2}]_J$ subshell configurations that can contribute to the $J^{\pi}=3^-$ octupole excitation for a $N_{\nu} = N_{\pi}$ combination of the oscillator shell quantum numbers, with $\rho_1$ or $\rho_2$ either being proton ($\pi$) or neutron ($\nu$), respectively. These proton-neutron configurations are realised for $N_n = Z$ nuclei near $A\approx 68$ and $112$. In the middle column the possible $J$ values are given and right-hand side examples for the oscillator shells $N_{\nu} = N_{\pi} =4$.}
\begin{indented}
\item[]\begin{tabular}{@{}ccc}
\br
$pn$-configuration & $|j_1-j_2|\leq J \leq j_1 +j_2$& Example\\
\mr
$[\pi 1(N+1)_{N+3/2}, \nu 1N_{N-1/2}]_J$ & $2 \leq J \leq 2N+1$ & $[\pi 1h_{11/2^-}, \nu 1g_{7/2^+}]$\\
&&\\
$[\pi 1N_{N-1/2}, \nu 1(N+1)_{N+3/2}]_J$ & $2 \leq J \leq 2N+1$ & $[\pi 1g_{7/2^+}, \nu 1h_{11/2^-}]$\\
&&\\
$[\pi 1(N+1)_{N+3/2}, \nu 2(N-2)_{N-3/2}]_J$ & $3 \leq J \leq 2N$ & $[\pi 1h_{11/2^-}, \nu 2d_{5/2^+}]$\\
&&\\
$[\pi 1(N-2)_{N-3/2}, \nu 1(N+1)_{N+3/2}]_J$ & $3 \leq J \leq 2N$ & $[\pi 2d_{5/2^+}, \nu 1h_{11/2^-}]$\\
\br
\end{tabular}
\end{indented}
\end{table}

For the quadrupole degree of freedom, it has been shown that the number of proton-neutron interactions, described by the $P$ factor:

\begin{equation}
P = \frac{n_{n} \cdot n_{p}}{n_{n} + n_{p}},
\end{equation}

\noindent which includes the number of valence neutron particles/holes $n_{n}$ and valence proton particles/holes $n_{p}$, governs the evolution of quadrupole collectivity \cite{Casten1,Casten2,Casten3}. This is reflected in the continuous drop of the excitation energy $E_{2^+_1}$ of the first $2^+_1$ level as a function of $P$. A plot of the excitation energy $E_{3^-_1}$ of the first $3^-_1$ state as a function of the $P$ factor is shown in Figs.~\ref{figure_6a} and \ref{figure_6b} for the $A\approx 144$ and $A\approx 222$ mass regions, respectively. Instead of being a monotonically decreasing function, the experimental excitation energies $E_{3^-_1}$ exhibit a minimum near $P$~factors that correspond valence nucleon numbers at and slightly beyond the octupole soft numbers. While the $3^-_1$ level for a few nuclei beyond the octupole-soft neutron numbers remains low, a slight increasing trend towards higher $P$ values is recognisable. If the proton-neutron interactions dominated, the plot would reveal a continuously decreasing $E_{3^-_1}$. While the proton-neutron interactions are certainly important, the $E_{3^-_1}$ excitation energy is mostly determined by other factors. Hence, this contribution will focus on the particle-particle and particle-hole couplings for likewise particles. Furthermore, since the valence shell subshell combinations contributing to the octupole phonon are realised northeast of doubly-magic nuclei, this work neglects the hole-hole channel as encountered southwest of doubly-magic nuclei. The combination of shell structure and Pauli blocking renders this channel less important when considering nuclei with enhanced octupole correlations.

\begin{table}[b]
\caption{\label{table_pnN1Z} Examples of proton-neutron $[\pi (n_1 l_1)_{j_1}, \nu (n_2 l_2)_{j_2}]_J$ subshell configurations that can contribute to the $J^{\pi}=3^-$ octupole excitation for a $N_{\nu} = N_{\pi} +1$ combination of the oscillator shell quantum numbers, with $\pi$ representing the protons and $\nu$ the neutrons, respectively. These proton-neutron configurations are realised for nuclei near $A\approx 90, 144,$ and $222$. In the middle column the possible $J$ values are given and the third column gives examples for $N_{\nu}= 5 = N_{\pi} +1$. Please note, other subshell combinations that can couple to $J^{\pi}=3^-$, but which do not exclusively contain the three subshells in the focus of this study, are not presented.}
\begin{indented}
\item[]\begin{tabular}{@{}ccc}
\br
$\pi\nu$-configuration & $|j_1-j_2|\leq J \leq j_1 +j_2$& Example\\
\mr
$[\pi 1(N+1)_{N+3/2}, \nu 1(N+2)_{N+5/2}]_J$ & $1 \leq J \leq 2N+4$ & $[\pi 1h_{11/2^-}, \nu 1i_{13/2^+}]$\\
&&\\
$[\pi 1N_{N-1/2}, \nu 1(N+1)_{N+1/2}]_J$ & $1 \leq J \leq 2N$ & $[\pi 1g_{7/2^+}, \nu 1h_{9/2^-}]$\\
&&\\
$[\pi 2(N-2)_{N-3/2}, \nu 2(N-1)_{N-1/2}]_J$ & $1 \leq J \leq 2N-2$ & $[\pi 2d_{5/2^+}, \nu 2f_{7/2^-}]$\\
&&\\
$[\pi 1N_{N-1/2}, \nu 2(N-1)_{N-1/2}]_J$ & $0 \leq J \leq 2N-1$ & $[\pi 1g_{7/2^+}, \nu 2f_{7/2^-}]$ \\
&&\\
$[\pi 2(N-2)_{N-3/2}, \nu 1(N+1)_{N+1/2}]_J$ & $2 \leq J \leq 2N-1$ & $[\pi 2d_{5/2^+}, \nu 1h_{9/2^-}]$\\
\br
\end{tabular}
\end{indented}
\end{table}

It is the aim of this contribution to investigate the $J^-$ level sequences for the previously introduced two-subshell combinations for the particle-particle and particle-hole cases. This work focuses on valence-shell negative-parity excitations that can couple to $J^{\pi}=3^-$. These are only a subset of two-subshell combinations that were investigated in Ref.~\cite{Schiffer76} in a more general approach with calculations using more sophisticated interactions. However, to maintain a maximum level of clarity, this work uses the most simplistic approach possible. The starting point of the description is the unperturbed energy of the two-particle system $E[(n_1 l_1)_{j_1}, (n_2 l_2)_{j_2};J]$. Without considering any residual interaction, $E[(n_1 l_1)_{j_1}, (n_2 l_2)_{j_2};J]$ is given as the energy difference between the two subshells and the pairing energy. To extract these energy differences from data obtained in particle-transfer experiments and information about the pairing from mass-measurement data is a tedious and often rather ambiguous task (e.g., see Ref.~\cite{ET}). The effect of shrinking subshell energy differences due to changes to the potential well with increasing mass number, which is obvious from the global behaviour of the excitation energies $E_{3^-_1}$ (e.g., see Fig.~1(a) in Ref.~\cite{Kibedi}), is neglected in this work. This work aims at the local minima of the $E_{3^-_1}$ data visible just above magic numbers. For simplicity, in this work, the unperturbed particle-particle (particle-hole) excitation energy $E[(n_1 l_1)_{j_1}, (n_2 l_2)_{j_2};J]=0$ is set to be zero.  The matrix elements provide the energy shifts $\Delta E[(n_1 l_1)_{j_1}, (n_2 l_2)_{j_2};J]$ caused by a perturbative interaction $\hat{V}_{res}$:

\begin{equation}
\Delta E[(n_1 l_1)_{j_1}, (n_2 l_2)_{j_2};J] = \langle (n_1 l_1)_{j_1}, (n_2 l_2)_{j_2}; J | \hat{V}_{res} | (n_1 l_1)_{j_1}, (n_2 l_2)_{j_2}; J \rangle.
\end{equation}

\noindent In the interest of simplicity, the two-body matrix elements are calculated using harmonic oscillator wavefunctions and a residual $\delta$ interaction. The possible $J$ values that can be realised for a two-subshell combination are given by the triangle inequality

\begin{equation}
\label{CSineq}
|j_1 - j_2| \leq J \leq |j_1 + j_2|,
\end{equation}

\noindent defined by the coupled angular momentum $j_i = l_i \pm 1/2$ of particles $i=1$ and $2$. 

In the following, the approach as outlined in References~\cite{Casten1,Talmi,Schiffer76,Heyde,Suhonen} is used.  For two likewise particles in non-equivalent orbits, a condition that is for the subshell combinations considered in this contribution always fulfilled, the energy shift $\Delta E[(n_1 l_1)_{j_1}, (n_2 l_2)_{j_2};J]$ can be decomposed

\begin{equation}
\label{ME_T1}
\Delta E[(n_1 l_1)_{j_1}, (n_2 l_2)_{j_2};J] = -V_{S=0} \cdot F_R[(n_1 l_1), (n_2 l_2)] \cdot A[(j_1), (j_2); J]
\end{equation}

\noindent into an empirical strength constant $V_{S=0}$, the Slater integral of the radial parts of the wavefunctions $F_R[(n_1 l_1), (n_2 l_2)]$, and an angular part $A[(j_1), (j_2); J]$. 

For a residual interaction in the form of a $\delta$ interaction, $\delta = \frac{1}{4\pi r^2}$, the Slater integral of the radial part calculates as \cite{Heyde}:

\begin{equation}
\label{radialoverlapp}
F_R[(n_1 l_1), (n_2 l_2)]=\frac{1}{4\pi}\int\limits_{0}^{\infty} \frac{1}{r^2} u_{n_1 l_1}^2(r) u_{n_2 l_2}^2(r) dr,
\end{equation}

\noindent where $u_{n_i l_i} = r \cdot R(n_i l_i)$ is related to the radial wavefunction $R(n_i l_i)$. The radial wavefunctions of the three-dimensional harmonic oscillator are given as \cite{Talmi}:

\begin{equation}
\label{oscilwf}
u_{nl}(r) = N_{n^{\prime}l} \cdot r^{l+1} e^{-b r^2} v_{n^{\prime}l}(2b r^2),
\end{equation}

\noindent where $n^{\prime} = n-1$. The normalisation factor $N_{n^{\prime}l}$ is defined by:

\begin{equation}
N_{n^{\prime}l} = \sqrt{\frac{2^{l-n^{\prime}+2}(2b)^{l+3/2}\cdot (2l+2n^{\prime}+1)!!}{\sqrt{\pi}[(2l+1)!!]^2 n^{\prime}!}}
\end{equation}

\noindent and the function $v_{n^{\prime}l}(2br^2)$, which is an alternative formulation for the Laguerre polynomials, is given as

\begin{equation}
v_{n^{\prime}l}(2b r^2) = \sum\limits_{k=0}^{n^{\prime}}(-1)^k 2^k \left( \begin{array}{c} n^{\prime}\\ k \end{array}\right) \frac{(2l+1)!!}{(2l+2k+1)!!} (2b r^2)^k.
\end{equation}

\noindent The inverse oscillator length $b=\frac{m\omega}{2\hbar}$ contains the oscillation frequency $\omega$ and the nucleon mass $m$ and can be approximated \cite{Heyde} by

\begin{equation}
\label{osclength}
b = \frac{41 A^{-1/3} mc^2}{2(\hbar c)^2} \left[\frac{1}{\textrm{fm}^2} \right]. 
\end{equation}

\begin{figure}[ht]
\centering
\includegraphics[width=0.7\textwidth, viewport=1 1 395 158, clip=true]{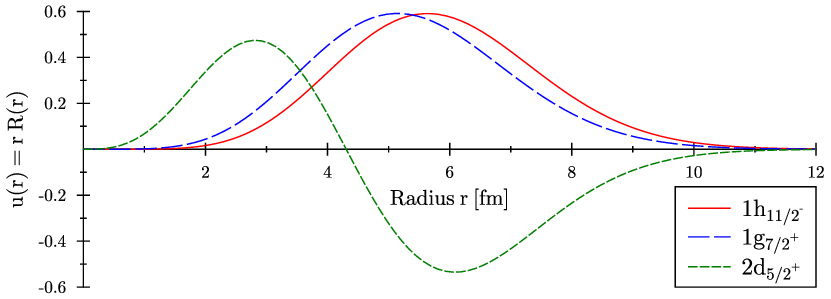}
\caption{Renormalised radial wavefunction $u_{nl} = r \cdot R_{nl}$ as function of the radius $r$. The $u_{nl}$ functions according to Eq.~(\ref{radialWFs}) are presented for the subshells in the $N=4$ oscillator shell that can couple to a $J^{\pi}=3^-$ state.}
\label{figure_7}
\end{figure}

\noindent Examples for $u(n_i l_i)$ corresponding to the $N=4$ oscillator shell are shown in Fig.~\ref{figure_7}. Considering that the principal quantum number $n$ of the involved subshells corresponds to the number of nodes of the radial wavefunction, where one root is found at $r=0$, it is evident that the $nl_j=2(N-2)_{N-3/2}$ subshell entering the $\Delta j =3, \Delta l=3$ configuration has, unlike the other two wavefunctions, a node at $r \neq 0$. The radial wavefunctions entering the $\Delta j =3, \Delta l =1$ subshell configuration both have exclusively a node at $r=0$. Hence, the overlap of the radial wavefunctions of the latter configuration and, therefore, $F_R((n_1 l_1), (n_2 l_2))$ can be expected to be larger than for the $\Delta j =3, \Delta l=3$ configuration.

Furthermore, the subshell combinations for protons and neutrons are realised in different mass regions. For example, the $N=4: [2d_{5/2^+},1h_{11/2^-}]$ subshell combination is realised for the neutrons near $A\approx 90$ and $112$, while it is realised for the protons near $A\approx 112$ and $144$. Therefore, the inverse oscillator length $b$ and, consequently, the Slater integrals take different values. Numerical values for $b$ calculated using Eq.~(\ref{osclength}) are presented in Column~3 of Table~\ref{table_4}. 

Within this approach, the radial parts of the three relevant wavefunctions $u_{nl}= u_{2(N-2)}, u_{1N}$, and $u_{1(N+1)}$ parametrised for the oscillator quantum number $N$ are given as:

\begin{equation}
\label{radialWFs}
\begin{array}{rcl}
u_{2(N-2)} &=& \sqrt{\frac{2^{N-1}(2b)^{N-1/2}}{\sqrt{\pi} (2N-1)!!}} r^{N-1} e^{-b r^2} (2N -1 - 4b r^2)\\
&&\\
u_{1N} &=& \sqrt{\frac{2^{N+2} (2b)^{N+3/2}}{\sqrt{\pi} (2N+1)!!}} r^{N+1} e^{-b r^2}\\
&&\\
u_{1(N+1)} &=& \sqrt{\frac{2^{N+3} \cdot (2b)^{N+5/2}}{\sqrt{\pi} \cdot (2N+3)!!}} r^{N+2} e^{-b r^2}.
\end{array}
\end{equation}

\noindent The calculation of the Slater integral requires the integral-identity \cite{Bronstein}:

\begin{equation}
\int\limits_{0}^{\infty} x^n e^{-\alpha x^2} dx = \frac{(n-1)!! \sqrt{\pi}}{2^{[(n+2)/2]} \alpha^{[(n+1)/2]}}, 
\end{equation}

\noindent which in this form is valid for even values of $n$. Since the wavefunctions enter quadratically, the condition of an even $n$ in the integrand is always fulfilled. However, the reader should bear in mind that for odd values of $n$ the solution of the integral takes a different form \cite{Bronstein}. Interestingly, since the Slater integral does not contain a $J$ dependency, it provides only a scaling factor for the energy splitting of the $J$ states of a two-subshell coupling in a given mass region, in addition to the empirical strength constant $V_{S=0}$. 

From the previous consideration, it is the angular part $A[(j_1),(j_2); J]$ that determines the $J$-dependent energy shift $\Delta E[(n_1 l_1), (n_2 l_2; J]$. It can be expressed as:

\begin{equation}
\label{Angularpart}
A[(j_1), (j_2); J] = \left\{\begin{array}{cccc}
(2j_1+1)(2j_2+1)\left(\begin{array}{ccc}j_1 & j_2 & J\\1/2 & -1/2 & 0\end{array}\right)^2 & \textrm{for} & l_1 + l_2 + J &\textrm{even}\\
&&&\\
0 & \textrm{for} & l_1 + l_2 + J & \textrm{odd.}\end{array} \right.\ 
\end{equation}

\noindent The sum $l_1 + l_2$ can, in terms of the oscillator quantum number $N$, be written  as $l_1 = N+1, l_2 =N \Rightarrow l_1+l_2 = 2N+1$  for the $\Delta j=3, \Delta l=1$ configuration and $l_1 = N+1, l_2 = N-2 \Rightarrow  l_1+l_2 = 2N-1$ for the $\Delta j=3, \Delta l=3$ configuration. Consequently, both $l_1 +l_2$ combinations are, for any given $N$, odd. Hence, a coupling to an even angular momentum $J$ will result in $A[(j_1),(j_2); J]=0$ and, therefore, $\Delta E[(n_1 l_1)_{j_1}, (n_2 l_2)_{j_2};J]=0$. In practical terms, for a system of two likewise particles, the energy of an accumulation of non-collective even-$J^-$ levels indicates the energy of an unperturbed particle-particle excitation.

\section{particle-particle channel}

\subsection{$\Delta j=3, \Delta l = 3$ configuration}

As previously stated, this particle-particle excitation contains the $\pi_1 = (-1)^{N+1}$ unique parity subshell $(n_1 l_1)_{j_1} =  1(N+1)_{N+3/2}$ and the $\pi_2 = (-1)^N$ subshell $(n_2 l_2)_{j_2} = 2(N-2)_{N-3/2}$. For this subshell combination, the overlap of the Slater integral $F_R[2(N-2), 1(N+1)]$ can be expressed as:

\begin{equation}
F_R[2(N-2), 1(N+1)] = \left(\frac{b}{\pi}\right)^{3/2} \cdot \frac{1}{2^{2N}} \cdot \frac{(4N-1)!}{(2N+1)!! \cdot (2N-1)!!}.
\end{equation}

\noindent The corresponding numerical values for the oscillator shells $N$ found in the mass regions $A$ are given in Column~5 of Table~\ref{table_4}. The decreasing overlap with increasing mass scales with the increasing volume within which the particles are distributed for the growing subshells.   


\begin{table}[b]
\caption{\label{table_4} The oscillator shell quantum number $N$, the mass number $A$, the resulting oscillator length $b$, and the Slater integral $F_R[(n_1 l_1), (n_2 l_2)]$ of the radial wavefunctions calculated using the wavefunctions of the three-dimensional harmonic oscillator provided by Eq.~(\ref{radialWFs}). For a discussion see text.}
\begin{indented}
\item[]\begin{tabular}{@{}ccccc}
\br
$N$ & $A$ & $b$ & $F_R[1N, 1(N+1)] \times 10^3$ & $F_R[2(N-2), 1(N+1)] \times 10^3$\\
&&[1/fm$^2$]&[1/fm$^3$]&[1/fm$^3$]\\
\mr
3 & 68 & 0.121 & 1.21 & 0.78\\
&&&&\\
3 & 90 & 0.110 & 1.05 & 0.68\\
&&&&\\
4 & 90 & 0.110 & 0.86 & 0.53\\ 
&&&&\\
4 & 112 & 0.103 & 0.77 & 0.47\\
&&&&\\
4 & 144 & 0.094 & 0.68 & 0.42\\
&&&&\\
5 & 144 & 0.094 & 0.57 & 0.34\\
&&&&\\
5 & 222 & 0.082 & 0.46 & 0.27\\
&&&&\\
6 & 222 & 0.082 & 0.40 & 0.23\\
\br
\end{tabular}
\end{indented}
\end{table}

Using $j_1 = N+1+1/2 = N+3/2$ and $j_2 = N-2+1/2 = N-3/2$ to express Eq.~(\ref{Angularpart}) in terms of the oscillator-shell quantum number $N$ results for levels with an odd-$J$ in: 

\begin{equation}
\label{ppdl3}
A[(N+3/2), (N-3/2); J] = 4(N-1)(N+2)\left(\begin{array}{ccc}N+3/2 & N-3/2 & J\\1/2 & -1/2 & 0\end{array}\right)^2.
\end{equation}

\begin{figure}[ht]
\centering
\includegraphics[width=0.7\textwidth, viewport=0 0 435 575, clip=true]{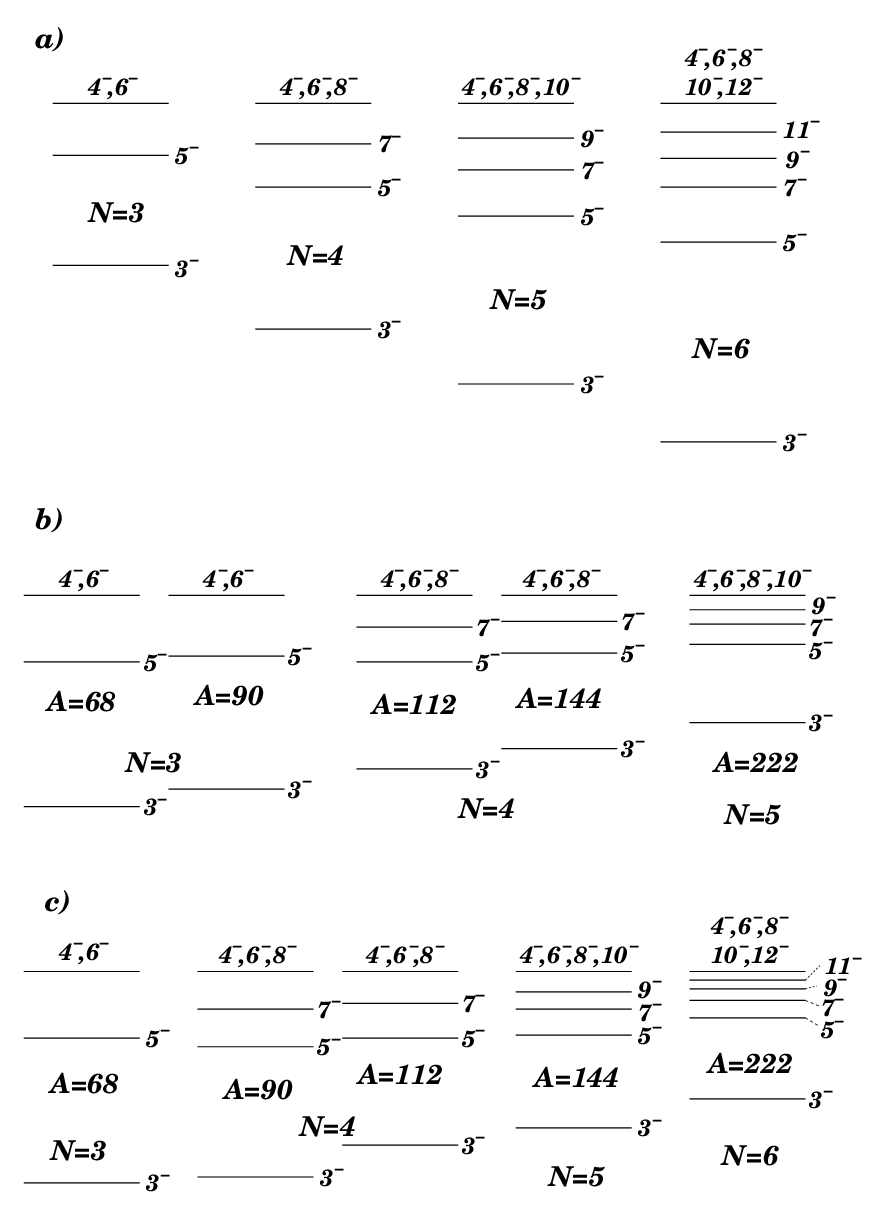}
\caption{\label{d3l3} Level sequences for the $\Delta j=3, \Delta l=3$ couplings in the particle-particle channel calculated using a $\delta$ interaction. Part a) presents the sequences as a function of the oscillator shell quantum number $N$ and only the angular part $A[(j_1), (j_2); J]$ is considered. The product of the Slater integral and the angular part $F_R[2(N-2), 1(N+1)]\cdot A[(j_1), (j_2); J]$ is plotted for the relevant two-proton configurations in part~b) and for the two-neutron configurations in  part~c).}
\label{figure2}
\end{figure}

\noindent According to Eq.~(\ref{CSineq}), the possible $J$ values lie in the range $3 \leq J \leq 2N$. For the odd-$J$ couplings, the values of these 3-$j$ symbols, as well as all other 3-$j$ and 6-$j$ symbols used in this work, were calculated using the online calculator of Ref.~\cite{Stone}. The $A[(N+3/2), (N-3/2); J]$ values are presented in Table~\ref{table_2} and the resulting level schemes are shown in Fig.~\ref{figure2}a). 

\begin{table}[b]
\caption{\label{table_2} Numerical values for the angular part $A_N[(j_1), (j_2); J]$ for the possible odd-$J$ values of a fixed oscillator quantum number $N$ of the $\Delta j=3, \Delta l=3$ subshell configuration.  Here $j_1 = N+3/2$ represents the intruder subshell and $j_2 = N-3/2$ the subshell with $\Delta j =3, \Delta l=3$ relative to the intruder subshell.}
\begin{tabular}{@{}ccccc}
\br
$J$ & $A_3[(9/2), (3/2); J]$ &   $A_4[(11/2), (5/2); J]$ & $A_5[(13/2), (7/2); J]$ & $A_6[(15/2), (9/2); J]$ \\
\mr
3 & $\frac{40}{21}$ & $\frac{200}{77}$ &  $\frac{2800}{858}$ & $\frac{1120}{286}$ \\
&&&&\\
5 & $\frac{40}{66}$ & $\frac{280}{286}$ &  $\frac{560}{429}$ & $\frac{7840}{4862}$\\
&&&&\\
7& & $\frac{1008}{2145}$ & $\frac{9072}{12155}$& $\frac{90720}{92378}$ \\
&&&&\\
9 & & &  $\frac{1680}{4199}$ & $\frac{5280}{8398}$ \\
&&&&\\
11&&&&  $\frac{5280}{14858}$ \\
\br
\end{tabular}
\end{table}

In Table~\ref{table_2} and Fig.~\ref{d3l3}a), the tendency that $A[(N+3/2), (N-3/2); J]$ is largest for the lowest $J$ value and its decrease with increasing $J$ is recognisable. The lowest-$J$ corresponds to a configuration in which the two involved spins are antialigned. Clearly, the $\Delta j =3, \Delta l=3$ coupling provides a considerable energy gain for the $J^{\pi} =3^-$ state. Interestingly, the angular part gains more energy with increasing oscillator quantum number $N$ [Part~a)], but as shown in Part~b) and Part~c) of Fig.~\ref{d3l3}, the product $F_R[(n_1 l_1), (n_2 l_2)]\cdot A[(j_1), (j_2); J]$ of both the Slater integral and the angular parts results in a decreasing energy shift $\Delta E[(n_1 l_1)_{j_1}, (n_2 l_2)_{j_2}]$ with increasing oscillator quantum number $N$ and mass number $A$.

\subsection{$\Delta j=3, \Delta l = 1$ configuration}

As previously stated, this particle-particle excitation contains the unique parity subshell $(n_1 l_1)_{j_1} = 1(N+1)_{N+3/2}$ and the  subshell $(n_2 l_2)_{j_2} = 1N_{N-1/2}$. For this subshell configuration, the Slater integral $F_R[1N, 1(N+1)]$ can be expressed as:

\begin{equation}
F_R[1N, 1(N+1)] =  \left(\frac{b}{\pi}\right)^{3/2} \cdot \frac{1}{2^{2N+1}} \cdot \frac{(4N+3)!}{(2N+1)!! \cdot (2N+3)!!}.
\end{equation}

\noindent The numerical values of the Slater integral are for the various oscillator shells $N$ as realised in the mass region $A$ given in Column~4 of Table~\ref{table_4}. 

Expressing Eq.~(\ref{Angularpart}) in terms of the oscillator-shell quantum number $N$  with $j_1 = N+3/2$ and $j_2 = N-1/2$ results for levels with an odd-$J$ in: 

\begin{equation}
\label{ppdl1}
A[(j_1), (j_2); J] = 4N(N+2)\left(\begin{array}{ccc}N+3/2 & N-1/2 & J\\1/2 & -1/2 & 0\end{array}\right)^2.
\end{equation}

\begin{figure}[ht]
\centering
\includegraphics[width=0.7\textwidth, viewport=0 0 435 575, clip=true]{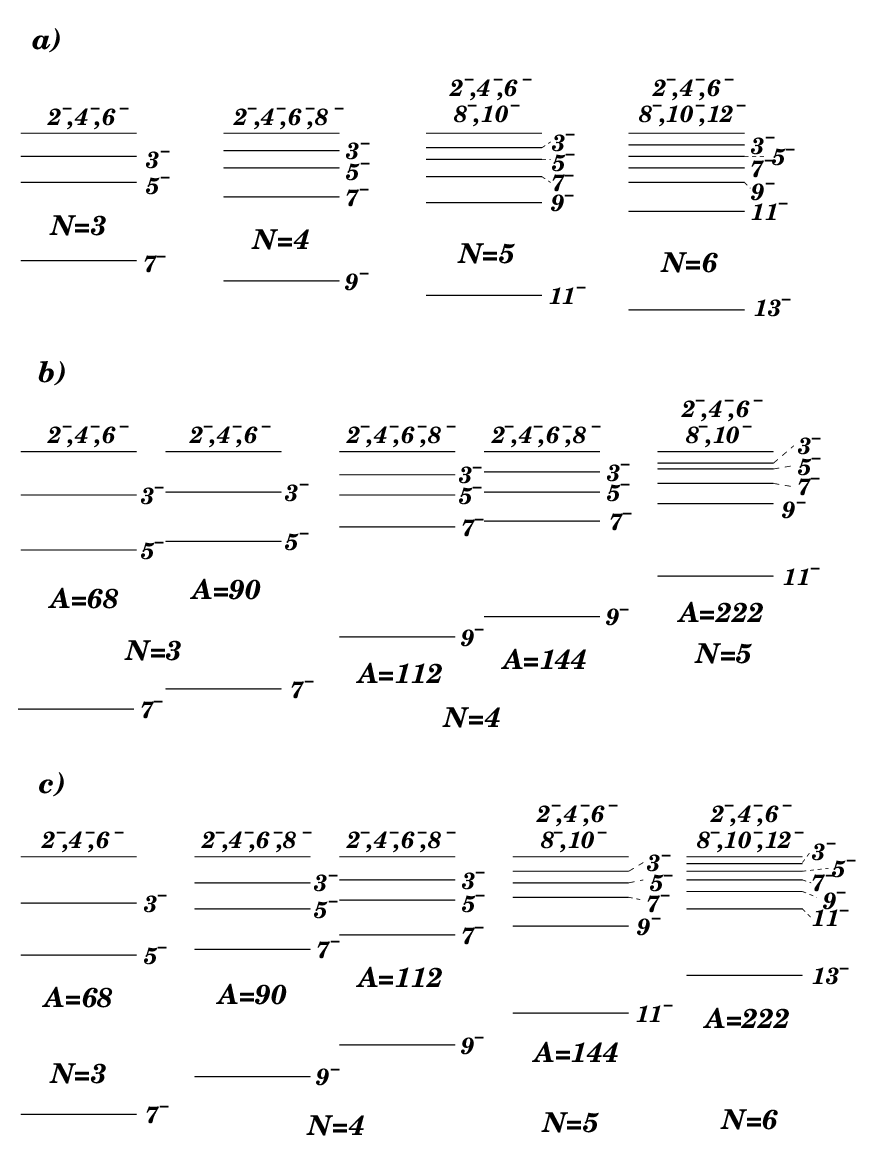}
\caption{\label{d3l1} Level sequences for the $\Delta j=3, \Delta l=1$ couplings in the particle-particle channel calculated using a $\delta$ interaction. In part~a) only the angular part $A[(j_1), (j_2); J]$ with $j_1 = N+3/2$ and $j_2 = N-1/2$ is considered. The product of Slater integral and angular parts $F_R(1(N), 1(N+1))\cdot A[(j_1), (j_2); J]$ is plotted for the two-proton configurations [Part~b)] and for the two-neutron configurations [Part~c)] as realised for the oscillator shell $N$ in the mass region $A$. Please note, the same scaling factors as used in Figure~\ref{figure2} were applied for all parts.}
\end{figure}

\noindent According to Eq.~(\ref{CSineq}), the $J$ values lie in the range $2 \leq J \leq 2N+1$. The numerical values of $A[(j_1), (j_2); J]$ are given in Table~\ref{table_3} and the resulting level sequences are shown in Fig.~\ref{d3l1}. 

\begin{table}[b]
\caption{\label{table_3} Results for the angular part $A_N[(j_1), (j_2); J]$ for the possible odd-$J$ values of a fixed oscillator quantum number $N$ for the $\Delta l=3, \Delta j=1$ configuration. Here, $j_1 = N+3/2$ represents the intruder subshell and $j_2 = N-1/2$ the subshell with $\Delta j =3, \Delta l=1$.}
\begin{tabular}{@{}ccccc}
\br
$J$ & $A_3[(9/2), (5/2); J]$ &  $A_4[(11/2), (7/2); J]$ &  $A_5[(13/2), (9/2); J)$ &  $A_6[(15/2); (11/2); J]$ \\
\mr
3 & $\frac{60}{231}$ & $\frac{2400}{12012}$ & $\frac{140}{858}$ & $\frac{1344}{9724}$ \\
&&&&\\
5 & $\frac{240}{429}$ & $\frac{672}{1716}$ & $\frac{2240}{7293}$ & $\frac{47040}{184756}$\\
&&&&\\
7 & $\frac{420}{286}$ & $\frac{3360}{4862}$ & $\frac{22680}{46189}$ & $\frac{72000}{184756}$ \\
&&&&\\
9& & $\frac{14112}{8398}$ & $\frac{3360}{4199}$ & $\frac{221760}{386308}$\\
&&&&\\
11 & & & $\frac{13860}{7429}$ & $\frac{133056}{148580}$ \\
&&&&\\
13&&&& $\frac{302016}{148580}$ \\
\br 
\end{tabular}
\end{table}

Comparing Figs.~\ref{d3l3} and \ref{d3l1}, it is obvious that despite the absolutes of the Slater integrals being smaller, the angular parts of the $\Delta j = 3, \Delta l = 3$ coupling lead to a significant lowering of the $J^{\pi}= 3^-$ level, while for the  $\Delta j = 3, \Delta j = 1$ the lowering is negligible. In the particle-particle channel, the simplistic approach used in this section confirms the preference of the $\Delta j = 3, \Delta l = 3$ configuration.

\section{Particle-hole channel}

In nuclei for which the Fermi level is situated in a subshell above the $(n_1 l_1)_{j_1} = 2(N-2)_{N-3/2}$ or the $1N_{N-1/2}$ subshell, the bilinear form does contain a particle excitation into the unique parity subshell $(n_2 l_2)_{j_2} = 1(N+1)_{N+3/2}$ and a hole excitation in the other involved subshell.  

Particle-particle and particle-hole excitations are linked by the Pandya transformation \cite{Suhonen,Pandya}:

\begin{equation}
\label{Pandya_trafo_long}\begin{array}{l}
\Delta E[\left((n_1l_1)_{j_1}\right)^{-1}, (n_2 l_2)_{j_2}; J] = \\
\\
- \sum\limits_{J^{\prime}} (2J^{\prime}+1) \left\{\begin{array}{ccc} j_1& j_2 & J\\ j_1& j_2 & J^{\prime} \end{array}\right\} \Delta E[(n_1 l_1)_{j_1}, (n_2 l_2)_{j_2}; J^{\prime}]. 
\end{array}
\end{equation}


\noindent Here, $\Delta E[(n_1 l_1)_{j_1}, (n_2 l_2)_{j_2}; J^{\prime}]$ is the energy shift of the level with angular momentum $J^{\prime}$ of the particle-particle multiplet. This energy shift in the particle-particle channel is weighted with a prefactor including the 6-$j$~symbol $\{\cdots\}$. In the following, the reader should bear in mind that in the convention used in this work, a positive number corresponds to a lowering of the energy, while a negative number raises the energy of the particle-hole multiplet labeled by the angular momentum $J$.

\subsection{$\Delta j=3, \Delta l = 3$ configuration}

\begin{figure}[ht]
\centering
\includegraphics[width=0.7\textwidth, viewport=0 0 420 470, clip=true]{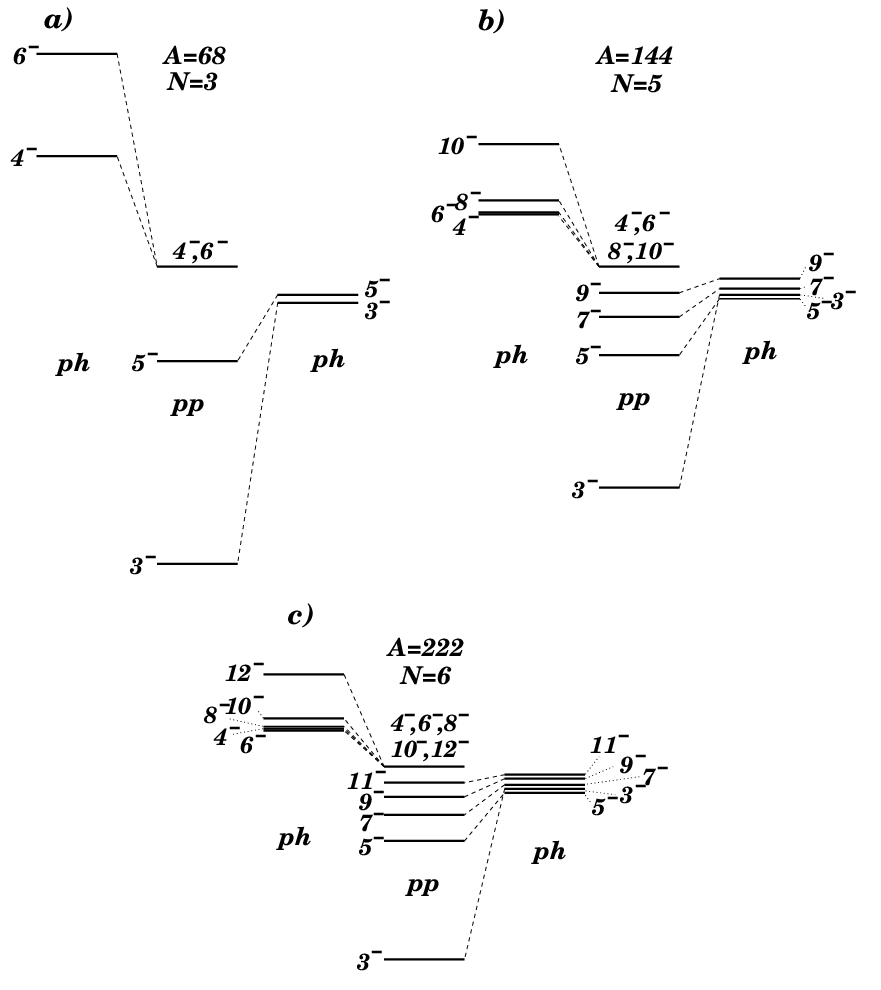}
\caption{Selected level sequences for the $\Delta j=3, \Delta l=3$ two-subshell couplings in the particle-hole channel (ph) are compared with those in the particle-particle channel (pp) as realised in the oscillator shell $N$ in the mass region $A$. The plots are organised in such a way that the even-$J$/odd-$J$ states in the ph channel are plotted on the left/right of the identical levels in the pp channel, which are presented in the center. The dashed lines are to guide the eye and connect the levels in pp and ph channels.}
\label{figure4}
\end{figure}

For the nuclei with $N \leq 6$, the $(n_1 l_1)_{j_1}= 2(N-2)_{N-3/2}$ subshell is settled below the $(n_2 l_2)_{j_2} = 1(N+1)_{N+3/2}$ intruder subshell. For heavier nuclei with $N>6$, this shell ordering might change due to the $l$-proportionality of the spin-orbit interaction, which causes the unique parity subshell to migrate deeper into the potential well with increasing orbital angular momentum. Anyhow, this work associates the hole excitation with the $2(N-2)_{N-3/2}$ subshell. The resulting energy shifts of the $J$ levels of the $\left[\left(2(N-2)_{N-3/2}\right)^{-1}, 1(N+1)_{N+3/2}\right]_J$ particle-hole excitation after applying the Pandya transformation are presented in Table~\ref{dl3ph} and, for selected multiplets, are shown in Fig.~\ref{figure4}. In the latter figure, the  level ordering of the particle-hole (ph) multiplet and the particle-particle (pp) channel are compared. In the particle-hole channel the degeneracy of the even-$J$ levels observed in the particle-particle channel is lifted and the even-$J$ levels are raised in energy. The energy loss is considerable for high-$J$ levels and decreases for lower-$J$ levels. The odd-$J$ levels gain some energy, but only a comparatively low amount with almost degenerated $J^{\pi}=3^-$ and $5^-$ levels and reduced energy gain with increasing $J$. Indeed, when compared with the particle-particle channel, the odd-$J$ levels are almost degenerate just below the unperturbed two-particle energy. Clearly, once the $nl_j= 2(N-2)_{N-3/2}$ subshell is filled and the Fermi level is found above this subshell, the energy gain due to the residual $\delta$ interaction is drastically reduced. Hence, the $\Delta j=3, \Delta l=3$ coupling is preferred in the particle-particle case, but to a lesser extent in the particle-hole channel. Therefore, octupole soft numbers can be expected before the $nl_j = 2(N-2)_{N-3/2}$ subshell is completly filled and one deals with a particle-partcle excitation rather than with a particle-hole configuration. Due to symmetry considerations, the results for a possible $\left[2(N-2)_{N-3/2},\left(1(N+1)_{N+3/2}\right)^{-1}\right]_J$ configuration, in which the hole excitation is in the $1(N+1)_{N+3/2}$ intruder shell, are identical to the presented particle-hole results. 



\begin{table}[b]
\caption{\label{dl3ph} Numerical results for the relative energy shift $\Delta E(N/A;J)/V_{S=0} = \Delta E\left[\left(2(N-2)_{N-3/2}\right)^{-1}, 1(N+1)_{N+3/2}; J\right]/V_{S=0}$ of the $\Delta l=3, \Delta j=3$ particle-hole multiplet as a function of the coupled angular momentum $J$ for the given oscillator quantum numbers $N$ as realised near the mass number $A$. Please note that in this work a positive value indicates a gain in binding energy. }
\begin{tabular}{@{}ccccccccc}
\br
$J$ & $\frac{E(3/68;J)}{V_{S=0}}$ & $\frac{E(3/90;J)}{V_{S=0}}$ & $\frac{E(4/90;J)}{V_{S=0}}$ & $\frac{E(4/112;J)}{V_{S=0}}$ & $\frac{E(4/144;J)}{V_{S=0}}$ & $\frac{E(5/144;J)}{V_{S=0}}$ & $\frac{E(5/222;J)}{V_{S=0}}$ & $\frac{E(6/222;J)}{V_{S=0}}$\\
\mr
3 &  0.186 &  0.162 &  0.170 &  0.153 &  0.135 &  0.138 &  0.111 &  0.113\\
&&&&&&&&\\
4 & -0.544 & -0.473 & -0.371 & -0.332 & -0.293 & -0.260 & -0.209 & -0.195\\
&&&&&&&&\\
5 &  0.167 &  0.144 &  0.180 &  0.161 &  0.142 &  0.155 &  0.125&  0.130\\
&&&&&&&&\\
6 & -1.064 & -0.925 & -0.462 & -0.405 & -0.357 & -0.272 & -0.219 & -0.189\\
&&&&&&&&\\
7 &        &        &  0.104 &  0.093 &  0.082 &  0.106 &  0.085 &  0.095\\
&&&&&&&&\\
8 &        &        & -0.848 & -0.760 & -0.670 & -0.327 & -0.263 & -0.198\\
&&&&&&&&\\
9 &        &        &        &        &        &  0.061 &  0.049 & 0.065\\
&&&&&&&&\\
10 &       &        &        &        &        & -0.616 & -0.496 & -0.242\\
&&&&&&&&\\
11 &       &        &&&&&& 0.038\\
&&&&&&&&\\
12 &       &        &&&&&& -0.460\\
\br 
\end{tabular}
\end{table}

\subsection{$\Delta j=3, \Delta l = 1$ configuration}

\begin{figure}[ht]
\centering
\includegraphics[width=0.7\textwidth, viewport=0 0 420 450, clip=true]{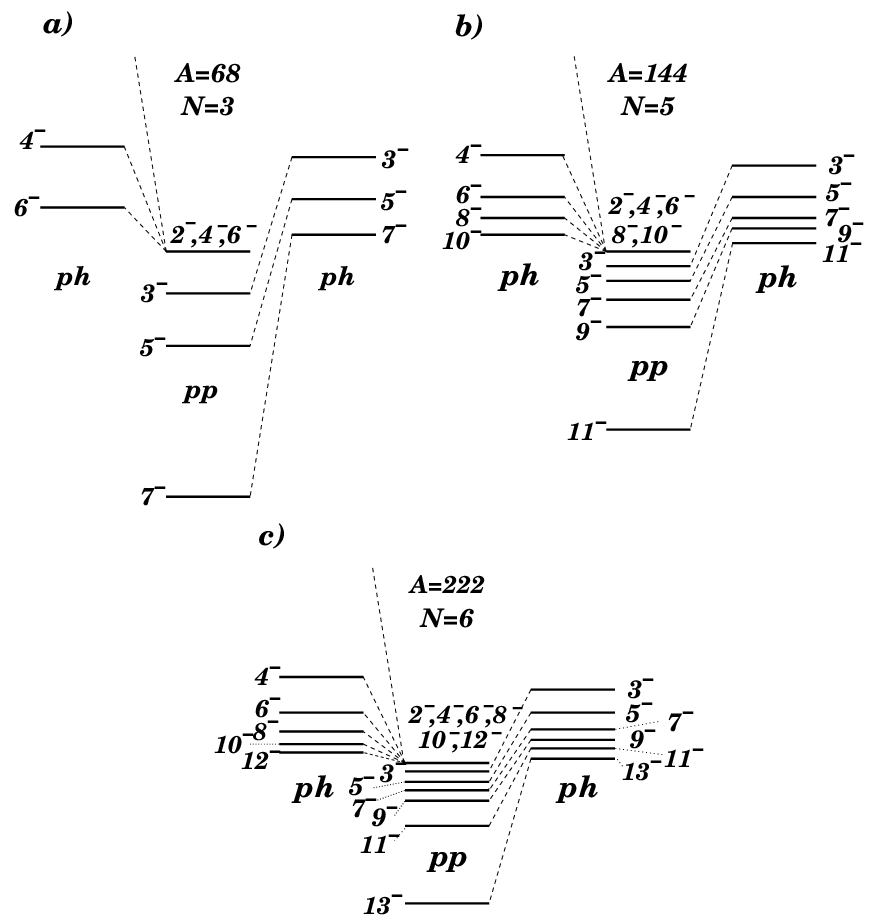}
\caption{Selected level sequences for the $\Delta j=3, \Delta l=1$ two-subshell couplings in the particle-hole channel (ph) compared with those in the particle-particle channel (pp) for the oscillator shell $N$ as realised near the mass number $A$. The plots are organised in such a way that the even-$J$/odd-$J$ states in the ph channel are plotted on the left/right of identical levels in the pp channel in the center. The dashed lines are to guide the eye and connect the levels in pp and ph channels. To maintain the clarity of the presentation for the other levels, the elevated position of the $2^-$ level is indicated but the level not included in the figure.}
\label{figure5}
\end{figure}

As in the previous subsection, the hole excitation will be associated with the $nl_j= 1N_{N-1/2}$ subshell originating from the $N$-th oscillator shell. As stated in the previous subsection, an inversion of particle and hole excitation results in identical energy shifts for the $J$ states of the particle-hole multiplet. The resulting energy shifts of the $J$ levels of the $\left[\left(1(N)_{N-1/2}\right)^{-1}, 1(N+1)_{N+3/2}\right]_J$ particle-hole excitation, after applying the Pandya transformation, are presented in Table~\ref{dl1ph} and, for selected multiplets, are shown in Fig.~\ref{figure5}.

\begin{table}[b]
\caption{\label{dl1ph} Results for the relative energy shift $E(N/A;J)/V_{S=0} = E\left[\left(1(N)_{N-1/2}\right)^{-1}, 1(N+1)_{N+3/2}; J\right]/V_{S=0}$ of the $\Delta l=1, \Delta j=3$ particle-hole multiplet as function of the coupled angular momentum $J$ for the given oscillator quantum numbers $N$ as realised near the mass number $A$. Please note that in this work a positive value indicates a gain in binding energy.}
\begin{tabular}{@{}ccccccccc}
\br
$J$ & $\frac{E(3/68;J)}{V_{S=0}}$ & $\frac{E(3/90;J)}{V_{S=0}}$ & $\frac{E(4/90;J)}{V_{S=0}}$ & $\frac{E(4/112;J)}{V_{S=0}}$ & $\frac{E(4/144;J)}{V_{S=0}}$ & $\frac{E(5/144;J)}{V_{S=0}}$ & $\frac{E(5/222;J)}{V_{S=0}}$ & $\frac{E(6/222;J)}{V_{S=0}}$\\
\mr
2 & -1.917 & -1.667 & -1.946 & -1.745 & -1.539 & -1.679 & -1.352 & -1.430\\
&&&&&&&&\\
3 & -0.453 & -0.394 & -0.471 & -0.422 & -0.372 & -0.410 & -0.331 & -0.352\\
&&&&&&&&\\
4 & -0.502 & -0.436 & -0.539 & -0.483 & -0.426 & -0.478 & -0.385 & -0.413\\
&&&&&&&&\\
5 & -0.255 & -0.222 & -0.293 & -0.263 & -0.232 & -0.267 & -0.215 & -0.234\\
&&&&&&&&\\
6 & -0.216 & -0.188 & -0.275 & -0.247 & -0.218 & -0.260 & -0.209 & -0.231\\
&&&&&&&&\\
7 & -0.085 & -0.073 & -0.174 & -0.156 & -0.138 & -0.176 & -0.142 & -0.162\\
&&&&&&&&\\
8 &        &        & -0.141 & -0.126 & -0.111 & -0.158 & -0.127 & -0.150\\
&&&&&&&&\\
9 &        &        & -0.060 & -0.054 & -0.047 & -0.110 & -0.088 & -0.112\\
&&&&&&&&\\
10 &       &        &        &        &        & -0.087 & -0.070 & -0.099\\
&&&&&&&&\\
11 &       &        &        &        &        & -0.039 & -0.031 & -0.072\\
&&&&&&&&\\
12 &       &        &&&&&& -0.057\\
&&&&&&&&\\
13 &       &        &&&&&& -0.026\\
\br 
\end{tabular}
\end{table}

In Fig.~\ref{figure5}, the resulting level sequences for the particle-hole channel, which are labeled as (ph), are compared to the multiplet sequence of the particle-particle channel. The latter are labeled as (pp) and found in the center. While the even-$J$ states are plotted to the left of the particle-particle levels, the odd-$J$ states are found to the right. The unperturbed energy is provided by the even-$J$ states of the particle-particle channel. For the even-$J$ states of the particle-hole excitation, the degeneracy observed in the particle-particle channel is lifted. When compared with the particle-hole $\Delta j=3, \Delta l=3$ configuration, the pattern is reversed and the $J^{\pi} = 2^-$ low-$J$ level experiences a pronounced shift towards lesser binding. Indeed, the resulting pattern for the even-$J$ states resembles an inverted seniority-two sequence with the particles situated in identical orbits. This loss in energy decreases with increasing $J$. Interestingly, the particle-hole channel preserves the feature of the particle-particle channel that the $J^{\pi} = 3^-$ level is found highest for the odd-$J$ states and the level energy is gradually lowered with increasing $J$. However, all levels experience a loss of binding when compared to the unperturbed energy. Consequently, the $\Delta j=3, \Delta l=1$ particle-hole configuration does not contribute to a lowering of the energy of the collective $3^-$ level and, therefore, represents the least favourable configuration to participate in the wavefunction. 





\section{Discussion}

The level sequences provided in the previous section validate, for the spherical case the statement that the $\Delta j=3, \Delta l=3$ two-subshell configuration is favoured over the $\Delta j=3, \Delta l=1$ two-subshell combination. The energy gain of low-$J$ states for the antialigned coupling of the angular momenta in the $\Delta j=3, \Delta l=3$ case and the preference of high-$J$ states in the aligned case of the $\Delta j=3, \Delta l=3$ combination reflect the tendency of the nuclear force to couple two likewise particles to spin $S=0$. Interestingly, this behaviour, that is due to the angular parts, overshadows for the $\Delta j= 3, \Delta l= 3$ configuration the less favourable overlap of the radial wavefunctions. Nevertheless, it must be repeated that the present results reflect pure configurations and are calculated within the simplistic approach of harmonic oscillator wavefunctions and a $\delta$-like residual interaction for the spherical case. When compared with experimental results, the calculated multiplet sequences provide a rough guideline for the underlying two-particle configuration. Anyhow, the enhanced energy gain of the $J^{\pi} = 3^-$ level for the $\Delta j=3, \Delta l=3$ two-subshell combination in the particle-particle channel explains why the octupole soft numbers occur two nucleons before the $nl_j = 2(N-2)_{N-3/2}$ subshell is fully occupied. Starting from two nucleons, with the increase of nucleons in the subshell the collectivity increases due to various interactions of the likewise particles, but also, even when neglected in this work, due to the proton-neutron interaction. Nevertheless, once the orbitals associated with the $2(N-2)_{N-3/2}$ subshell are filled, the particle-hole channel with its comparatively insignificant gain in energy dominates and the energy of the lowest-lying two-body $J^{\pi} =3^-$ configuration is comparatively high. As a consequence, the energy of the collective $3^-_1$ level is raised, too. Clearly, the admixture of valence-shell $3^-$ two-body excitations is most significant in the collective wavefunction and, therefore, their amplitudes in the collective wavefunction are considerable, while the amplitudes of cross-oscillator shell excitations may as a sum contribute strongly, but the amplitude of each individual component is miniscule. The number of cross oscillator shell two-body contributions that can couple to $J^{\pi}=3^-$ state increases with growing oscillator quantum number $N$. Furthermore, the gap between two oscillator shells decreases with increasing mass number and related width of the potential well. These two considerations provide a simple explanation for increased admixtures of these configurations to the octupole phonon. Clearly, these contributions are neglected in this work. 

\begin{figure}[ht]
\centering
\includegraphics[width=0.6\textwidth, viewport=0 0 400 330, clip=true]{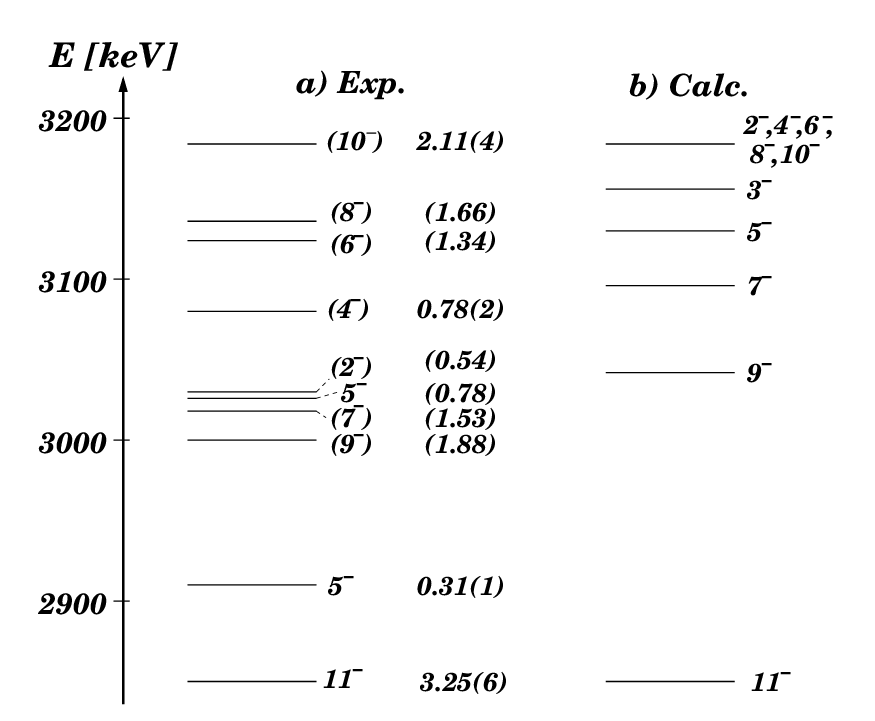}
\caption{Part a) shows candidate levels of the two-proton $[1h_{9/2},1i_{13/2}]_J$ multiplet as determined in the $^{209}$Bi$(\alpha,t)^{210}$Po reaction. Additionally, the relative spectroscopic factor as determined in this reaction is presented next to each level \cite{Groleau}. In part~b), the level sequence as calculated in this work is shown. For comparision the calculated energy shifts were normalised to the experimental energy difference between the $10^-$ and $11^-$ levels. For a discussion see text.}
\label{figure_8}
\end{figure}

In addition to the above mentioned limitations of the approach here, the reader must bear in mind that the present results are provided as a ratio $\Delta E/V_{S=0}$ of the energy shift $\Delta E$ normalised by the empirical strength constant $V_{S=0}$. The determination of the strength constant requires comparison to experiment. However, the experimental observation of comparatively pure negative-parity, valence-shell, seniority-two particle-particle or particle-hole multiplets is difficult. Starting from a suitable odd-mass target nucleus, one-particle transfer reactions like $(d,p)$, $(^{3}He,d)$, or $(\alpha, t)$ represent experimental probes to investigate the particle-particle multiplets. Ref.~\cite{Groleau} demonstrates that the use of the heavier $alpha$-particle is benefitial for the population of levels involving the unique partity with its high orbital angular momentum. A suitable target nucleus would be the neighboring odd-mass nucleus with the unpaired particle being either in the $2(N-2)_{N-3/2}$ or $1N_{N-1/2}$ subshells. By analogy, removal reactions like the $(p,d)$, $(d,^{3}He)$ or $(t,\alpha)$ are suitable tools to populate the particle-hole levels discussed in this work. Here a traget nucleus would have the Fermi level situated in the $1(N+1)_{N+3/2}$ unique parity subshell. Given that at least for the particle-particle $\Delta j=3, \Delta l=3$ multiplet a degeneracy of the even-$J$ levels is expected, a high energy resolution is required. Gates on $\gamma$~rays depopulating the excited levels would allow clean particle angular distributions to be determined. In order to minimise mixing with other close-lying levels of the same multipolarity, the level density in the investigated nucleus must be relatively low. Furthermore, one must consider that the multiplets under consideration can, as valence shell excitations only be observed in nuclei with $N\geq 3$. A condition that requires nuclei heavier than $^{56}$Ni. In order to keep level densities as low as possible, the use of semimagic nuclei is favourable. Furthermore, to keep the multiplet as pure as possible, the optimal choice is a nucleus with two valence nucleons added to a doubly-magic core. Unfortunately, the only stable, possible odd-mass target nucleus that fulfills all the above mentioned conditions is $^{209}$Bi with its $h_{9/2}$ ground state. Indeed, the $[1h_{9/2},1i_{13/2}]_J$ coupling in $^{210}$Po \cite{Schiffer76,Heyde} is so far the best example for a particle-particle multiplet as calculated in this contribution. A comparision of the experimentally observed levels to the results of the present calculations is shown in Fig.~\ref{figure_8}. In the figure, the data resulting from a $^{209}$Bi$(^{4}He,t)$ experiment \cite{Groleau} is compared to the calculations from the present work. In the experimental data there are deviations from those compiled in Ref.~\cite{Schiffer76}. For instance, the $3^-$ level is absent from the multiplet. Since it is the lowest-lying $3^-_1$ state, most likely admixtures of higher-lying $3^-$ states caused it to shift to lower energies and form the collective octupole phonon at 2386.7~keV \cite{Kibedi}, which is not observed in the $^{209}$Bi$(^{4}He,t)$ experiment \cite{Groleau}. The $5^-$ state is fragmented over two levels, with the lower-energy state attributed by Groleau and coworkers to include a core-excited configuration. This suggestion is supported by the lower relative spectroscopic factor determined in Ref.~\cite{Groleau}. These spectroscopic factors are included in Fig.~\ref{figure_8}~a) next to the level spin. If no uncertainty is given in the spectroscopic factors of Fig.~\ref{figure_8}, the corresponding peak was part of an unresolved multiplet in the particle spectrum. To compare the calculations (part~b of Fig.~\ref{figure_8}) the theoretical energies of the $J^{\pi} = 10^-$ and $11^-$ levels were adjusted to the experimental values. Since there are fewer configurations that can couple to these spins, these levels can be expect to be the purest in the multiplet. Indeed, the comparision reveals that the deviations between experiment and calculation are larger the lower the angular momentum of a level. This is true for both even-$J$ and odd-$J$ levels. Of course, the simplistic model approach used in this work neglects admixture of other configurations as well as more sophisticated interactions, such as the Coulomb-repulsion of the two protons \cite{Talmi}, which does lift the $J$ degeneracy of the even-$J$ levels \cite{Schiffer76}. 

Clearly, at present, there is a lack of experimental data. However, in the last two decades, the advent of radioactive ion beam facilities and the opportunity to post-accelerate intense beams of suitable nuclei to the required energies put transfer experiments in inverse kinematics within reach, especially near the doubly-magic nucleus $^{132}$Sn. Of course, starting from such nuclei with two particles outside a doubly-magic core, it will be very exciting to investigate the evolution of a multiplet along isotopic and isotonic chains. However, it is a tedious task to identify the levels for which the wavefunction has a particle-particle or particle-hole component with a sufficiently large amplitude . An example for such a work, where candidates for various two-body or even four-body excitations were identified, is provided by Ref.~\cite{GarrettZr}. In this publication, a multi-messenger approach is employed to characterise a large number of excited levels in the semi-magic nucleus $^{90}$Zr. 

\section{Summary}

This work used a simplistic approach to test the often-repeated claim that $\Delta j=3, \Delta l=3$ configurations are preferred over the $\Delta j=3, \Delta l=1$ two-body excitation in enhancing low-lying octupole collectivity. This work points out the reduced overlap of the radial parts of the wavefunction, but demonstrates that, for a $\delta$ interaction, the angular parts of the wavefunction more than compensate, resulting in a lowering of the odd-$J$ states. It is shown that this lowering is particularly pronounced in the particle-particle channel. A comparison to the sparse experimental data demonstrates the shortcomings but also the validity of the approach, serving at least as a starting point for a more sophisticated theoretical treatment. The simplicity of the model allows the reader to obtain a schematic approach to the valence-shell building blocks of the octupole excitation.

\section*{Acknowledgement}
This work is dedicated to the memory of the late Kris Heyde, who inspired it and supported it in its initial stage through countless advice. Furthermore, the authors wish to thank Juoni Suhonen for the generously given advice and the encouraging support. The authors gratefully acknowledge financial support by the UK-STFC (grant ST/P005101/1).

\end{document}